\documentclass[a4paper,11pt]{article}
\usepackage{pos}
\usepackage{xspace}
\usepackage{slashed}

\newcommand{\eqn}{equation}
\newcommand{\lam}{\lambda}
\newcommand{\lb}{\left(}
\newcommand{\rb}{\right)}
\newcommand{\TeV}{{\ensuremath\rm TeV}\xspace}
\newcommand{\GeV}{{\ensuremath\rm GeV}\xspace}
\newcommand{\fb}{{\ensuremath\rm fb}\xspace}

\newcommand{\be}{\beta}
\newcommand{\al}{\alpha}

\title{Constraining extended scalar sectors at current and future colliders}

\author*[a,b]{Tania Robens}

\affiliation[a]{Division of Theoretical Physics, Rudjer Boskovic Institute\\
  Bijenicka cesta 54, 10000 Zagreb, Croatia}

\affiliation[b]{Theoretical Physics Department, CERN\\ 1211 Geneva 23, Switzerland}

\emailAdd{trobens@irb.hr}

\abstract{In this proceeding, I summarize results on various new physics extensions of the Standard Model, for models with and without dark matter candidates. I discuss current constraints as well as rates and discovery prospects at future colliders.\\ RBI-ThPhys-2022-12, CERN-TH-2022-056}

\FullConference{%
  Corfu Summer Institute 2021 "School and Workshops on Elementary Particle Physics and Gravity"\\
  29 August - 9 October 2021\\
  Corfu, Greece
}

\begin{document}
\maketitle

\section{Introduction}

In this proceeding, I discuss various new physics models that extend the Standard Model (SM) by adding additional fields that transform as either singlets or doublets under the standard model gauge group. Most of the results presented here have already been discussed elsewhere, and I summarize our previous findings here. In particular, I discuss
\begin{itemize}
\item{}The real singlet extension of the SM, which comes with an additional scalar that transforms as a singlet. The model features one additional CP even neutral scalar. See \cite{Pruna:2013bma,Robens:2015gla,Robens:2016xkb,Ilnicka:2018def,DiMicco:2019ngk} for original literature as well as \cite{Robens:2021rkl} for results presented here;
\item{}The Inert Doublet Model (IDM) \cite{Deshpande:1977rw,Cao:2007rm,Barbieri:2006dq}, a two Higgs doublet model (THDM) with an additional $\mathbb{Z}_2$ symmetry that provides a dark matter (DM) candidate. Our original work can be found in \cite{Ilnicka:2015jba,Ilnicka:2018def,Kalinowski:2018ylg,Kalinowski:2020rmb};
\item{}The two real singlet extension (TRSM), where the SM scalar sector is extended by two additional gauge singlets, featuring in total three CP even neutral scalars that also allow for interesting cascade decays. The results discussed here have first been presented in \cite{Robens:2019kga,Papaefstathiou:2020lyp};
\item{}The THDMa, a two higgs doublet model that is enhanced by an additional pseudoscalar which serves as a portal to the dark matter sector. In the version of the model discussed here, the DM candidate is fermionic. See \cite{Ipek:2014gua,No:2015xqa,Goncalves:2016iyg,Bauer:2017ota,Tunney:2017yfp,LHCDarkMatterWorkingGroup:2018ufk,Robens:2021lov} for original work and \cite{Kalinowski:2022fot} for work containing the results discussed here.
\end{itemize}

All models are confronted with most recent theoretical and experimental constraints. Theory constraints include the minimization of the vacuum as well as the requirement of vacuum stability and positivity. We also apply constraints from perturbative unitarity and perturbativity of the couplings at the electroweak scale.

Experimental bounds include the agreement with current measurements of the properties of the 125 \GeV~ resonance discovered by the LHC experiments, as well as agreement with the null-results from searches for additional particles at current or past colliders. Furthermore, we impose constraints from electroweak precision observables (via $S,\,T,\,U$  parameters \cite{Altarelli:1990zd,Peskin:1990zt,Peskin:1991sw}), 
B-physics observables $\lb B\,\rightarrow\,X_s\,\gamma,\,B_s\,\rightarrow\,\mu^+\,\mu^-,\,\Delta M_s\rb$, and astrophysical observables (relic density and direct detection bounds). We use a combination of private and public tools in these analyses. In particular, we use HiggsBounds \cite{Bechtle:2020pkv},  HiggsSignals \cite{Bechtle:2020uwn}, 2HDMC \cite{Eriksson:2009ws}, SPheno \cite{Porod:2011nf}, Sarah \cite{Staub:2013tta}, micrOMEGAs \cite{Belanger:2018ccd,Belanger:2020gnr}, and MadDM \cite{Ambrogi:2018jqj}. Experimental numbers are taken from \cite{Baak:2014ora,Haller:2018nnx} for electroweak precision observables, \cite{combi} for $B_s\,\rightarrow\,\mu^+\,\mu^-$, \cite{Amhis:2019ckw} for $\Delta M_s$ and \cite{Planck:2018vyg} and  \cite{Aprile:2018dbl} for relic density and direct detection, respectively.
Bounds from $B\,\rightarrow\,X_s\gamma$ are implemented via a fit function from \cite{Misiak:2020vlo,mm}. Predictions for production cross sections shown here have been obtained using  Madgraph5 \cite{Alwall:2011uj}.

\section{Real singlet extension}
As a first simple example, we discuss a real singlet extension of the SM with a $\mathbb{Z}_2$ symmetry previously reported on in \cite{Pruna:2013bma,Robens:2015gla,Robens:2016xkb,deFlorian:2016spz,Ilnicka:2018def,DiMicco:2019ngk}. The $\mathbb{Z}_2$ symmetry is softly broken by a vacuum expectation value (vev) of the singlet field, inducing mixing between the gauge-eigenstates which introduces a mixing angle $\al$. The model has in total 5 free parameters. Two of these are fixed by the measurement of the $125\,\GeV$ resonance mass and electroweak precision observables. We then have
\begin{\eqn}
\sin\al,\,m_2,\,\tan\beta\,\equiv\,\frac{v}{v_s}
\end{\eqn}
as free parameters of the model, where $v\,(v_s)$ are the doublet and singlet vevs, respectively. We concentrate on the case where $m_2\,\geq\,125\,\GeV$, where SM decoupling corresponds to $\sin\al\,\rightarrow\,0$.

\begin{center}
\begin{figure}
\begin{center}
\begin{minipage}{0.48\textwidth}
\includegraphics[width=\textwidth]{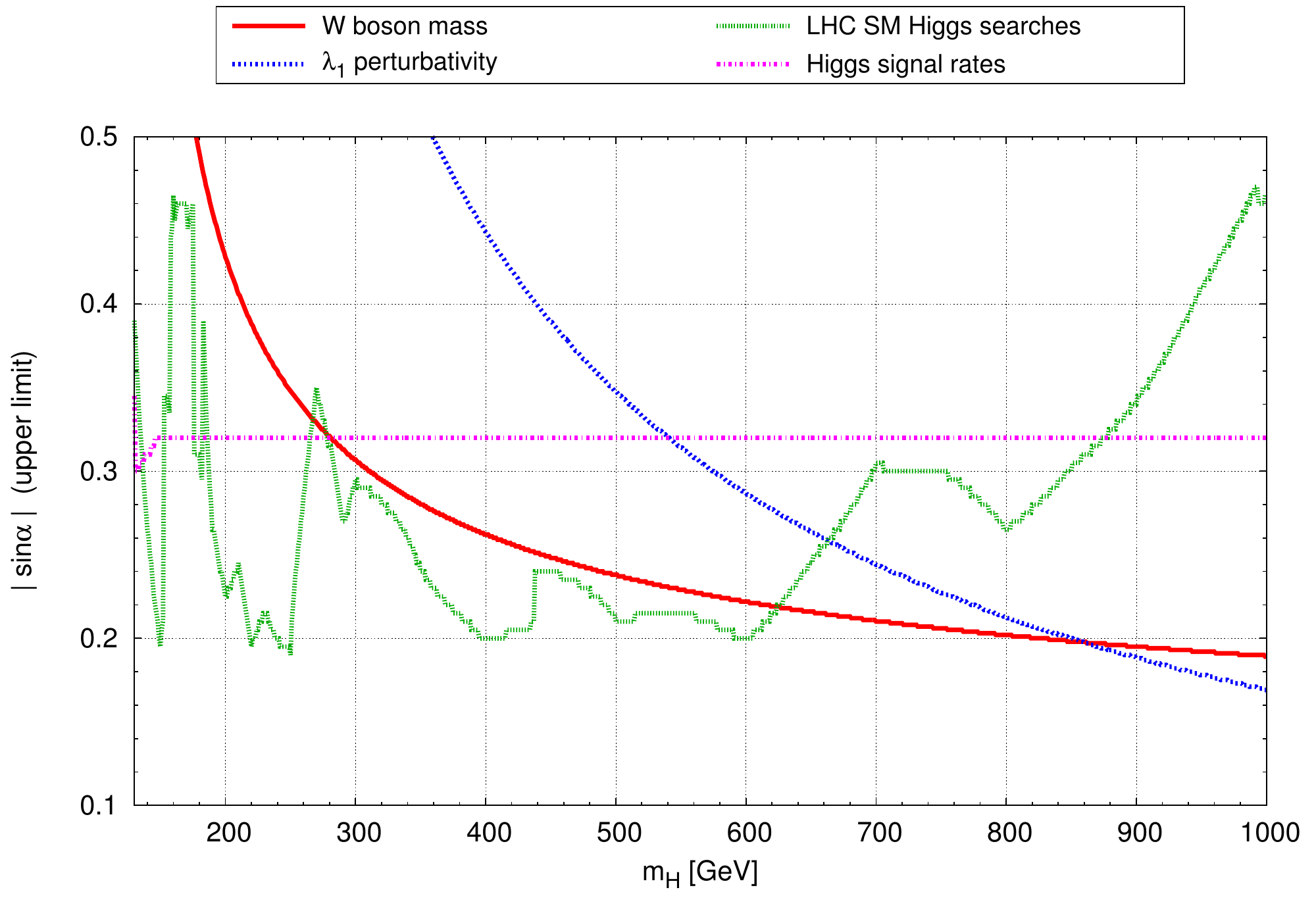}
\end{minipage}
\begin{minipage}{0.42\textwidth}
\includegraphics[width=\textwidth]{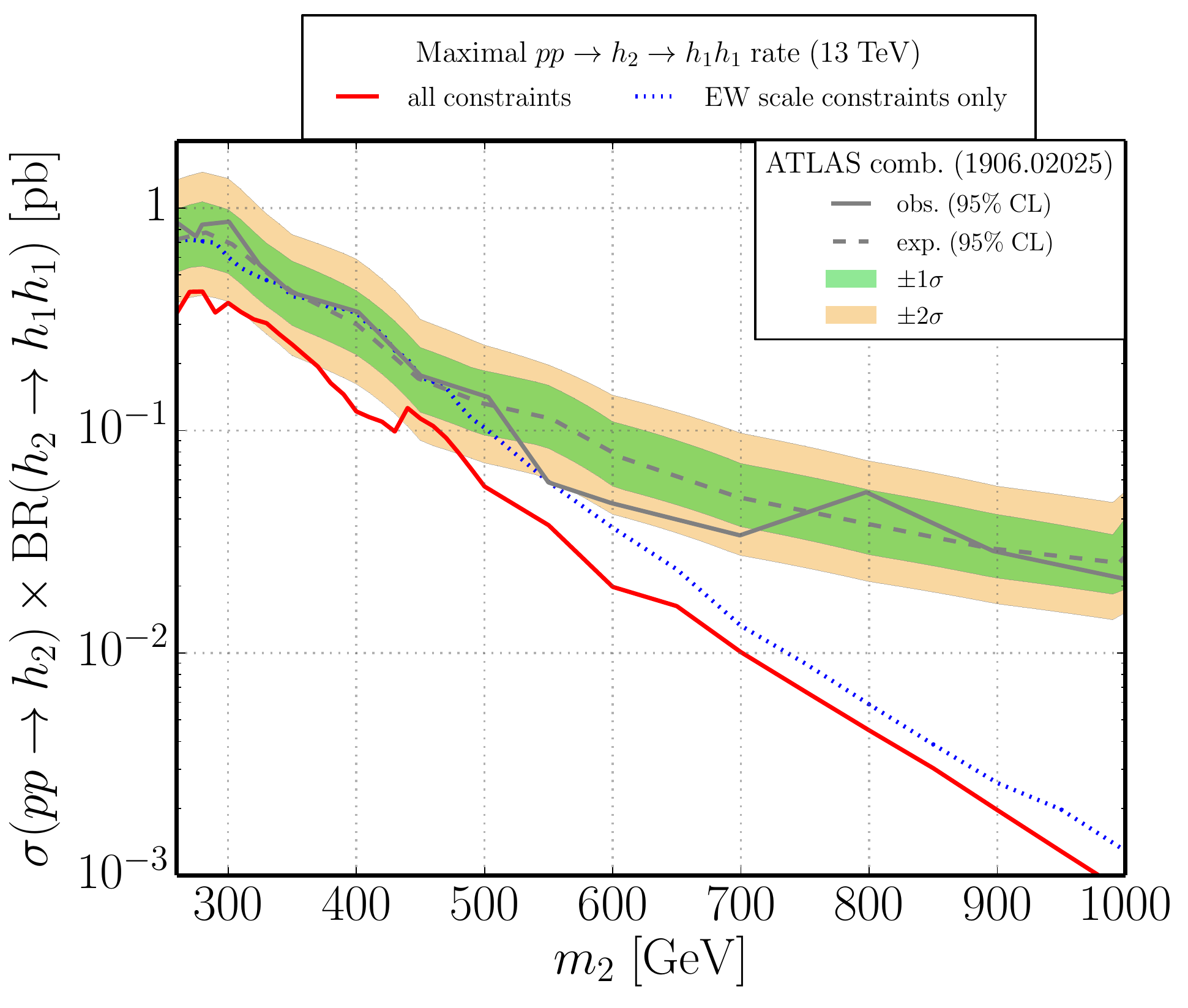}
\end{minipage}
\caption{Results for the singlet extension, taken from \cite{Robens:2021rkl}. {\sl Left:} comparison of current constraints for a fixed value of $\tan\beta\,=\,0.1$. {\sl Right:} maximal $H\,\rightarrow\,h\,h$ allowed, with electroweak constraints at the electroweak scale (blue) or including RGE running to a higher scale (red), in comparison with results from the ATLAS combination.}\label{fig:singlet}
\end{center}
\end{figure}
\end{center}

Limits on this model are shown in figure \ref{fig:singlet}, taken from \cite{Robens:2021rkl}\footnote{Updates will be presented in the proceedings of Moriond2022.}, including a comparison of the currently maximal available rate of $H\,\rightarrow\,h_{125}h_{125}$ with the combination limits from ATLAS \cite{Aad:2019uzh}. The most constraining direct search bounds are in general dominated by searches for diboson final states \cite{CMS-PAS-HIG-13-003,Khachatryan:2015cwa,Sirunyan:2018qlb,Aaboud:2018bun}. In some regions, the Run 1 Higgs combination \cite{CMS-PAS-HIG-12-045} is also important. Especially \cite{Sirunyan:2018qlb,Aaboud:2018bun} currently correspond to the best probes of the models parameter space\footnote{We include searches currently available via HiggsBounds.}.

\section{Inert Doublet Model}

\begin{figure}
\begin{minipage}{0.49\textwidth}
\begin{center}
\includegraphics[width=0.9\textwidth]{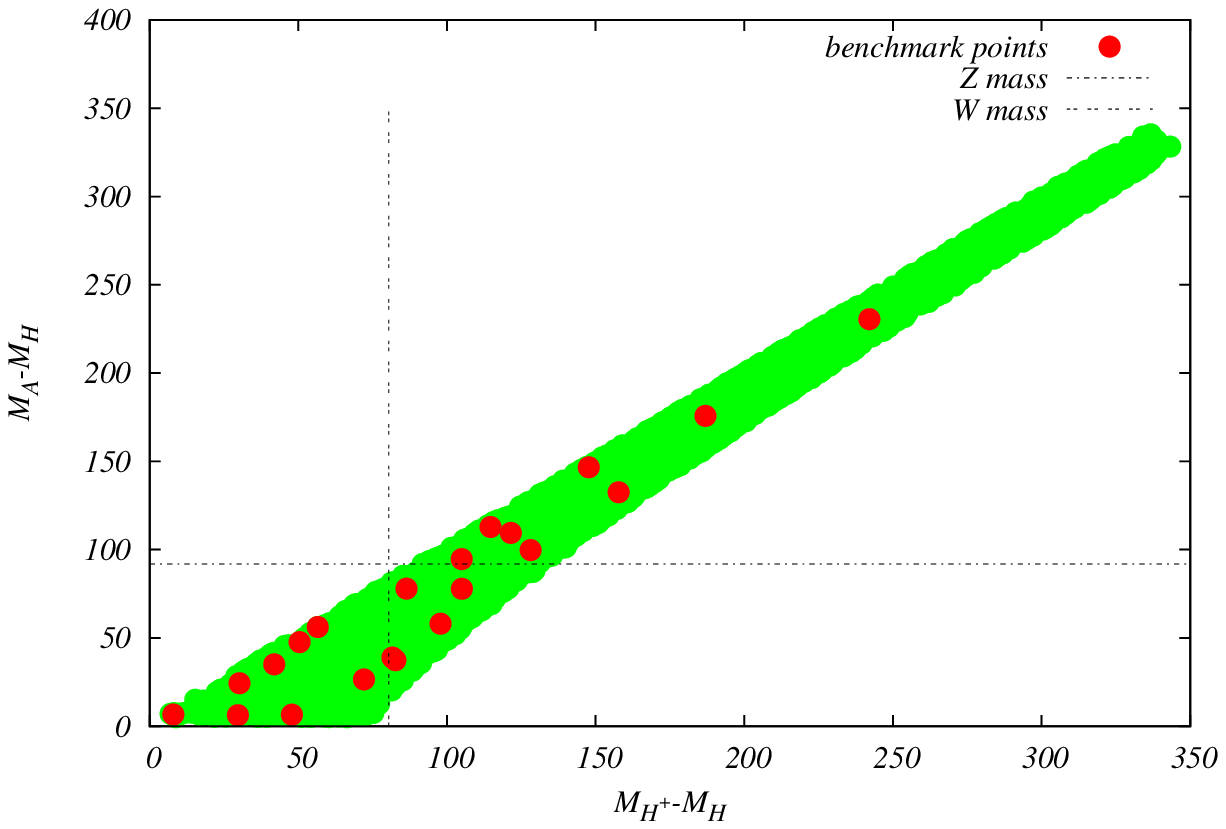}
\end{center}
\end{minipage}
\begin{minipage}{0.49\textwidth}
\begin{center}
\includegraphics[width=0.9\textwidth]{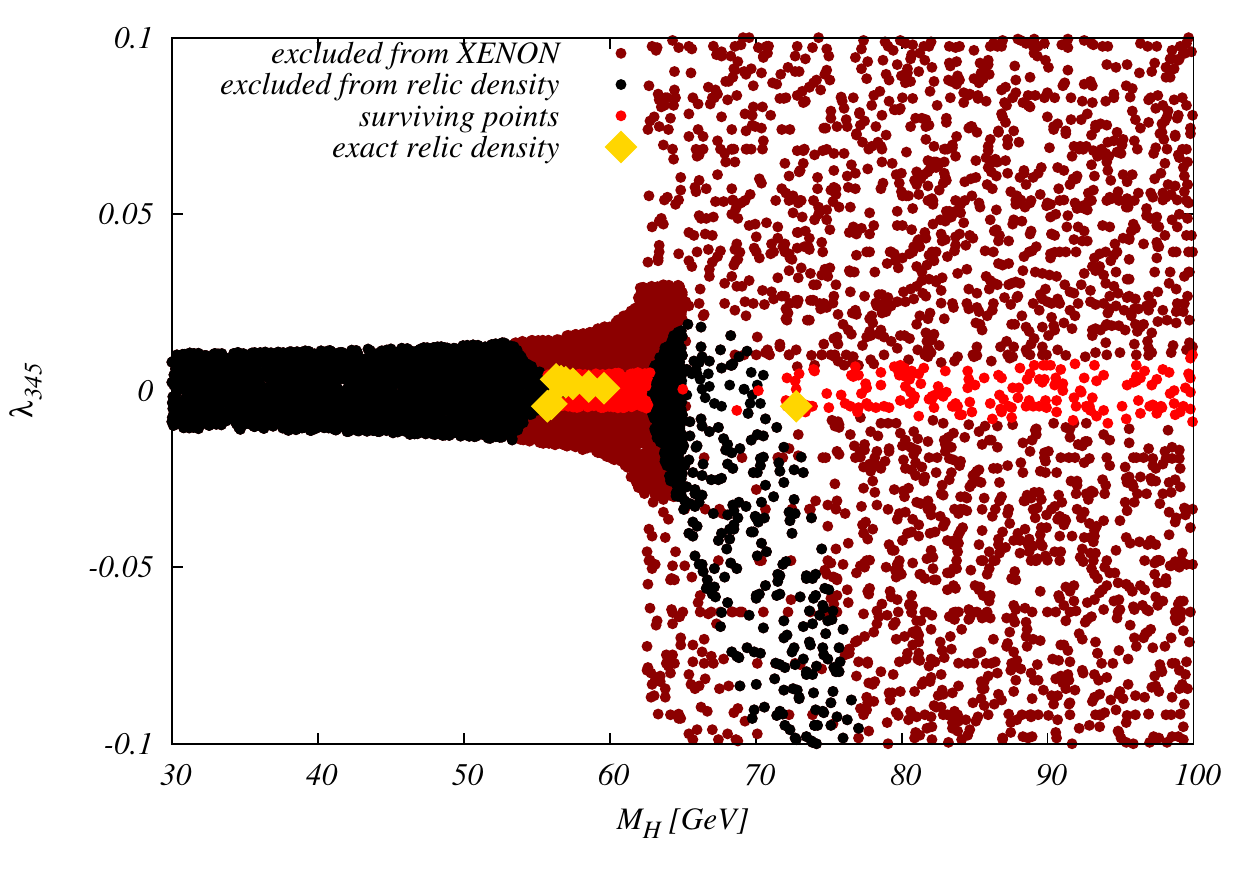}
\end{center}
\end{minipage}
\caption{\label{fig:idmscan} {\sl Left:} Masses are requested to be quite degenerate after all constraints have been taken into account. In the {$\lb M_{H^\pm}-M_H,\,M_A-M_H \rb$} plane (taken from \cite{Kalinowski:2018ylg}). {\sl Right:} Interplay of signal strength and relic density constraints in the $\lb M_H,\,\lam_{345}\rb$ plane, using XENON1T results, with golden points labelling those points that produce exact relic density (taken from \cite{Ilnicka:2018def}).}
\end{figure}

The Inert Doublet Model is a two Higgs doublet model with an exact discrete $\mathbb{Z}_2$ symmetry. It provides a dark matter candidate that stems from the second doublet \cite{Deshpande:1977rw,Cao:2007rm,Barbieri:2006dq}. The particle content of the model consists of four additional scalar states $H,\,A,\,H^\pm$, and has in total 7 free parameters prior to electroweak symmetry breaking:
\begin{\eqn}
v,\,m_h,\,\underbrace{m_H,\,m_A,\,m_{H^\pm}}_{\text{second doublet}},\,\lam_2,\,\lam_{345}\,\equiv\,\lam_3+\lam_4+\lam_5,
\end{\eqn}
Here, the $\lam_i$s denote standard couplings appearing in the THDM potential. Two parameters ($m_h$ and $v$) are fixed by current measurements. The model has been subjected to various experimental and theoretical constraints \cite{Ilnicka:2015jba,Ilnicka:2018def,Kalinowski:2018ylg,Kalinowski:2020rmb,Robens:2021yrl}. One important observation is the existance of a relatively strong degeneracy between the additional masses of the second doublet, as well as a minimal mass scale for the dark matter candidate $H$ resulting from a combination of relic density and signal strength measurement constraints (see \cite{Ilnicka:2015jba,Kalinowski:2020rmb} for a detailed discussion). We display these features in figure \ref{fig:idmscan}.

\subsection{Sensitivity study at current and future colliders}

\begin{figure}
\begin{center}
\begin{minipage}{0.42\textwidth}
\begin{center}
\includegraphics[width=\textwidth]{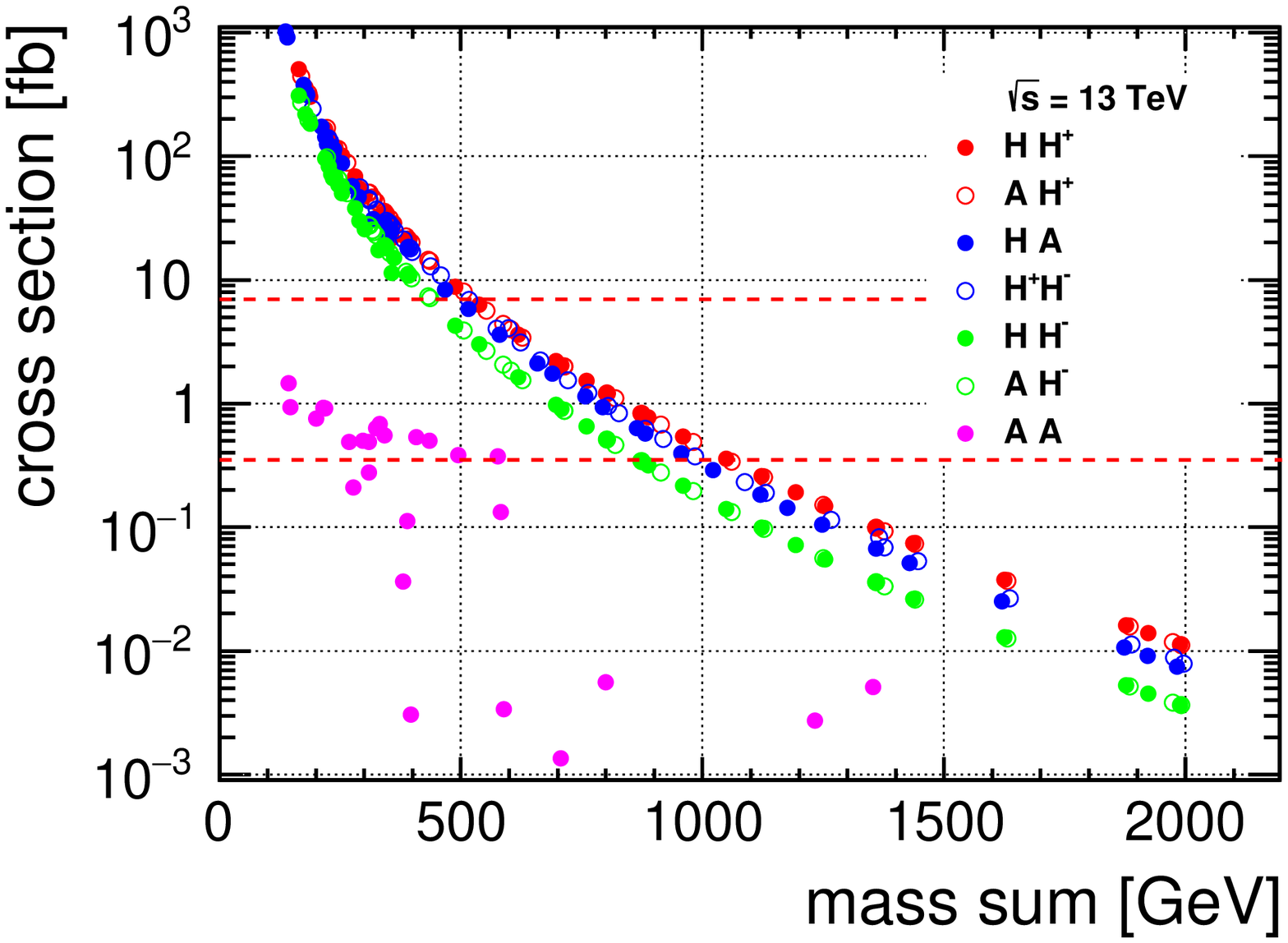}
\end{center}
\end{minipage}
\begin{minipage}{0.42\textwidth}
\begin{center}
\includegraphics[width=\textwidth]{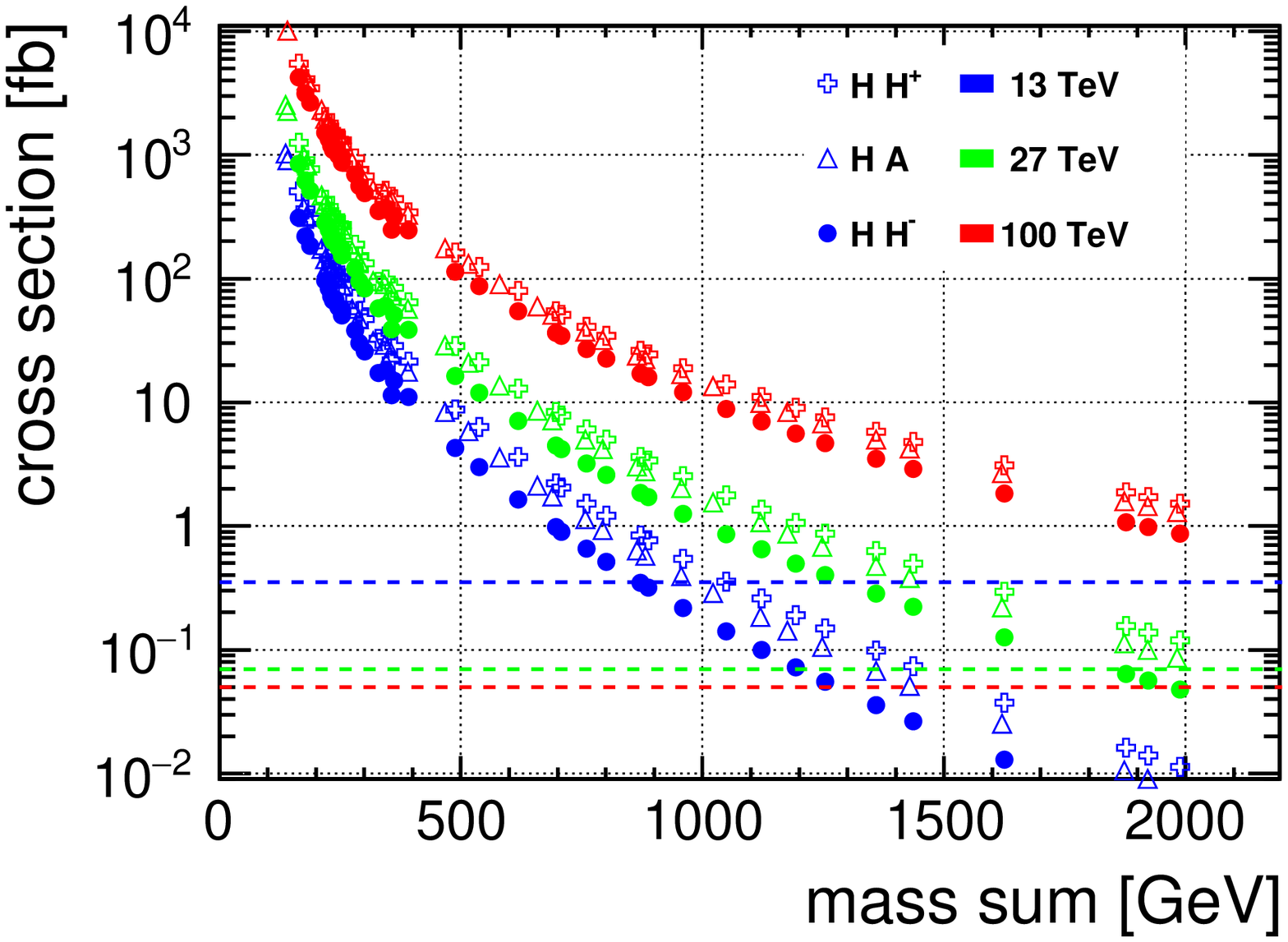}
\end{center}
\end{minipage}
\begin{minipage}{0.42\textwidth}
\begin{center}
\includegraphics[width=\textwidth]{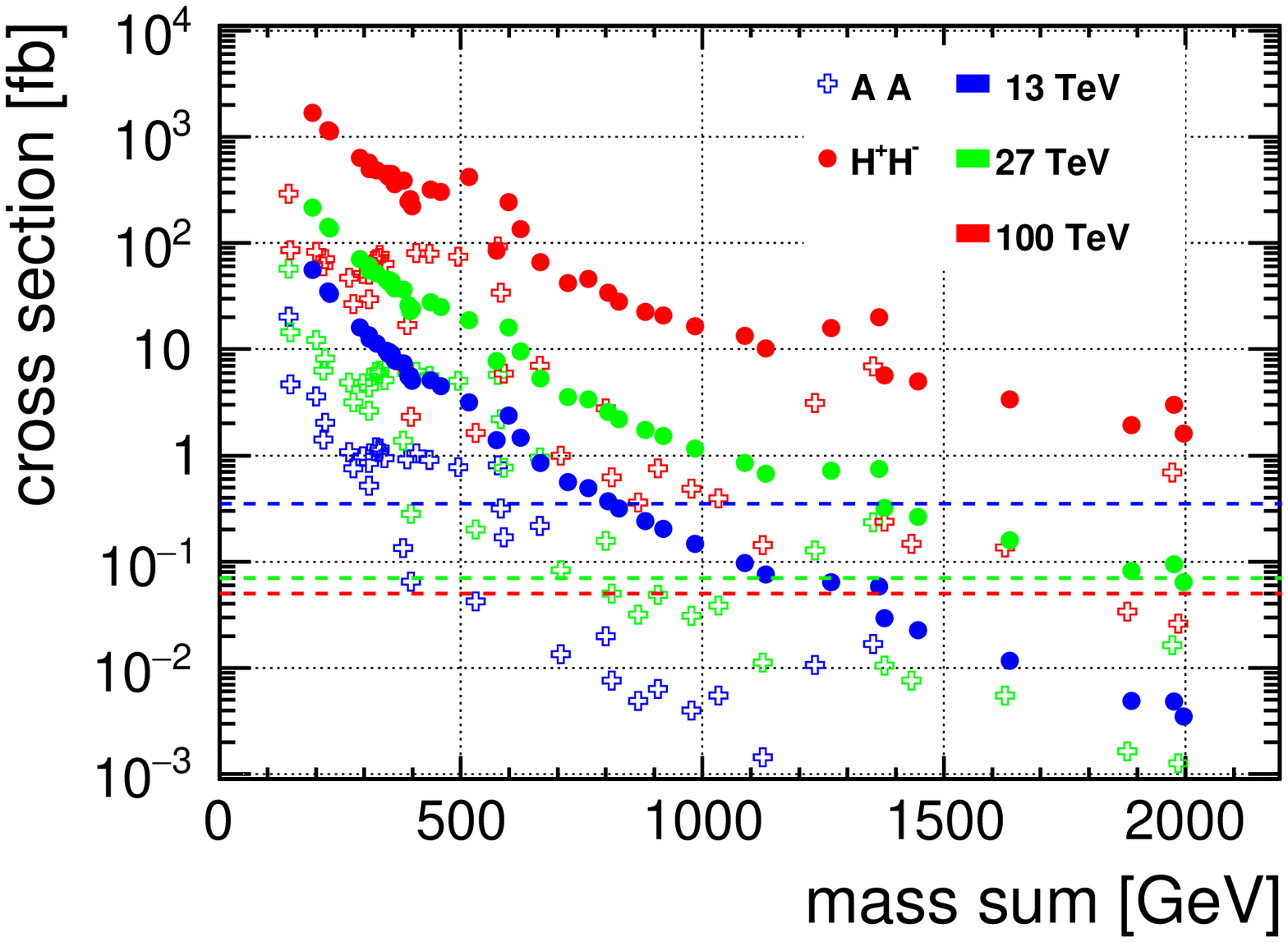}
\end{center}
\end{minipage}
\begin{minipage}{0.42\textwidth}
\begin{center}
\includegraphics[width=\textwidth]{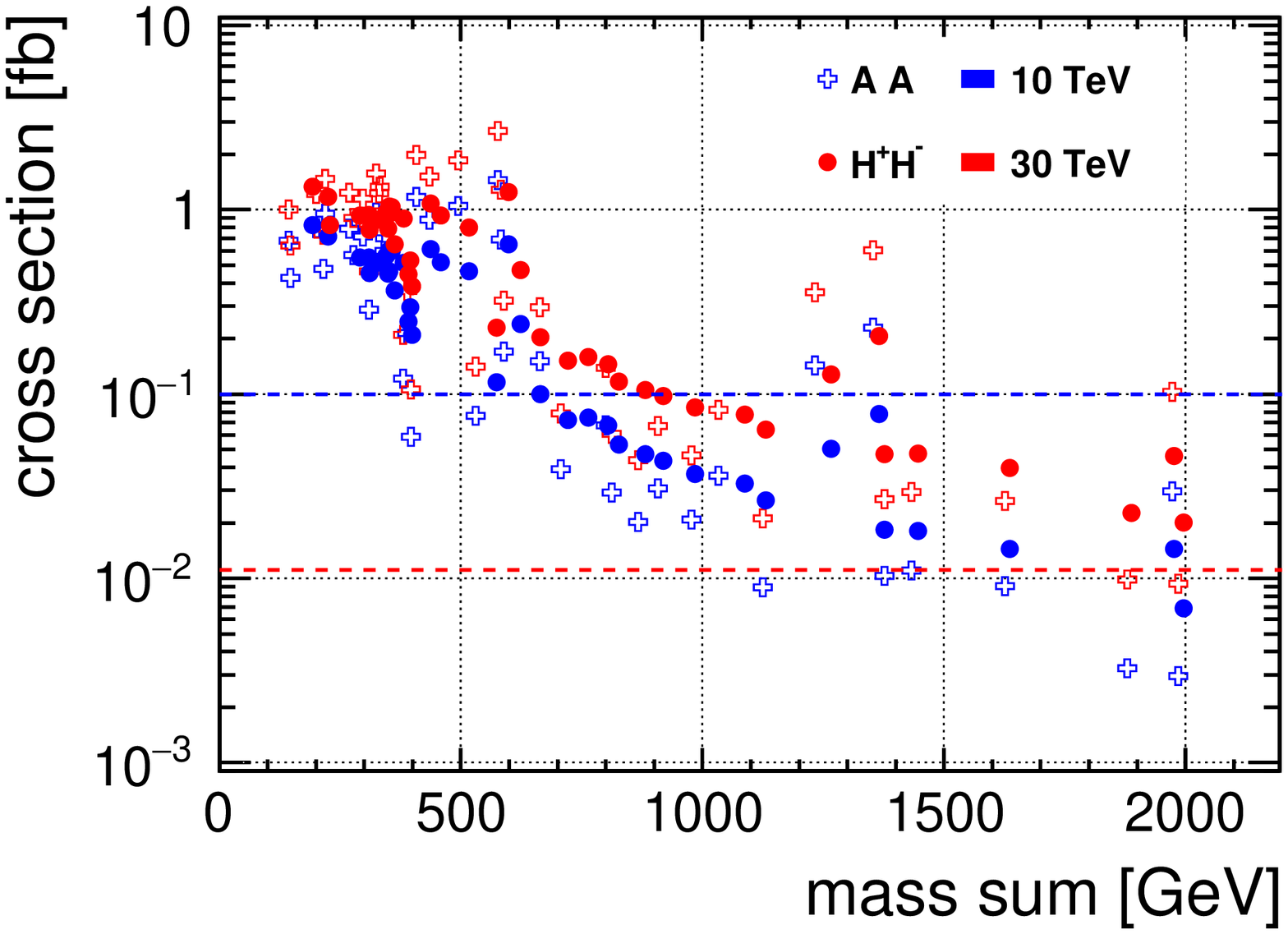}
\end{center}
\end{minipage}
\end{center}
\caption{\label{fig:idm} Predictions for production cross sections for various processes and collider options. {\sl Top left:} Predictions for various pair-production cross sections for a $pp$ collider at 13 \TeV, as a function of the mass sum of the produced particles. {\sl Top right:} Same for various center-of-mass energies. {\sl Bottom left:} VBF-type production of $AA$ and $H^+\,H^-$ at various center-of-mass energies for $pp$ colliders. {\sl Bottom right:} Same for $\mu^+\mu^-$ colliders. Taken from \cite{Kalinowski:2020rmb}. The lines correspond to the cross-sections required to prodce at least 1000 events using the respective design luminosity.}
\end{figure}

I here present results first discussed in \cite{Kalinowski:2020rmb}. In that work, a sensitivity comparison for selected benchmark points \cite{Kalinowski:2018ylg,Kalinowski:2018kdn,Kalinowski:2020rmb} was used, which relies on a simple counting criteria: a benchmark point is considered reachable if at least 1000 signal events are produced using nominal luminosity of the respective collider (c.f. also \cite{Robens:2021zvr}). Table \ref{tab:sens} shows the results using this simple criterium. The accompagnying figures, displaying production cross sections for pair-production of the novel scalars at various collider options and center-of-mass energies are shown in figure \ref{fig:idm}, taken from \cite{Kalinowski:2020rmb}. We here have used Madgraph5 \cite{Alwall:2011uj} with a UFO input file from \cite{Goudelis:2013uca} for cross-section predictions. Results for CLIC were taken from \cite{Kalinowski:2018kdn,deBlas:2018mhx}.

\begin{center}
\begin{table}
\begin{center}
\begin{tabular}{||c||c||c||c||} \hline \hline
{collider}&{all others}& { $AA$} & {$AA$ +VBF}\\ \hline \hline
HL-LHC&1 \TeV&200-600 \GeV& 500-600 \GeV\\
HE-LHC&2 \TeV&400-1400 \GeV&800-1400 \GeV\\
FCC-hh&2 \TeV&600-2000 \GeV&1600-2000 \GeV\\ \hline \hline
CLIC, 3 \TeV&2 \TeV &- &300-600 \GeV\\
$\mu\mu$, 10 \TeV&2 \TeV &-&400-1400 \GeV\\
$\mu\mu$, 30 \TeV&2 \TeV  &-&1800-2000 \GeV \\ \hline \hline
\end{tabular}
\end{center}
\caption{Sensitivity of different collider options, using the sensitivity criterium of 1000 generated events in the specific channel. $x-y$ denotes minimal/ maximal mass scales that are reachable.}
\label{tab:sens}
\end{table}
\end{center}
\section{TRSM}

In the TRSM \cite{Robens:2019kga}, the SM scalar sectors is augmented by two real scalars obeying a discrete $\mathbb{Z}_2\,\otimes\,\mathbb{Z}_2'$ symmetry. Both fields acquire a vev, which induces a mixing between all scalar states. The model then has 9 a priori parameters after electroweak symmetry breaking, 
\begin{\eqn*}
m_1,\,m_2,\,m_3,\,v,\,v_X,\,v_S,\,\theta_{hS},\,\theta_{hX},\,\theta_{SX},
\end{\eqn*}
where $m_i,\,v,\,\theta$ denote masses\footnote{We use the convention $m_1\,\leq\,m_2\,\leq\,m_3$.}, vevs, and mixing angles. One mass $m\,\sim\,125\,\GeV$ and $v\,\sim\,246\,\GeV$ are fixed by current measurements. 

Various benchmark planes (BPs) where proposed within this model \cite{Robens:2019kga}, allowing for novel production and decay processes, including decay chains which by that time had not been investigated by the LHC experiments. Production and decay modes can be characterized as 
\begin{\eqn*}
p\,p\,\rightarrow\,h_3\,\rightarrow\,h_1\,h_2,\;p\,p\,\rightarrow\,h_a\,\rightarrow\,h_b\,h_b,
\end{\eqn*}
where for the symmetric decays we assume none of the scalars corresponds to the SM-like 125 \GeV~ resonance.

\begin{center}
\begin{figure}
\begin{center}
 \includegraphics[width=0.55\columnwidth]{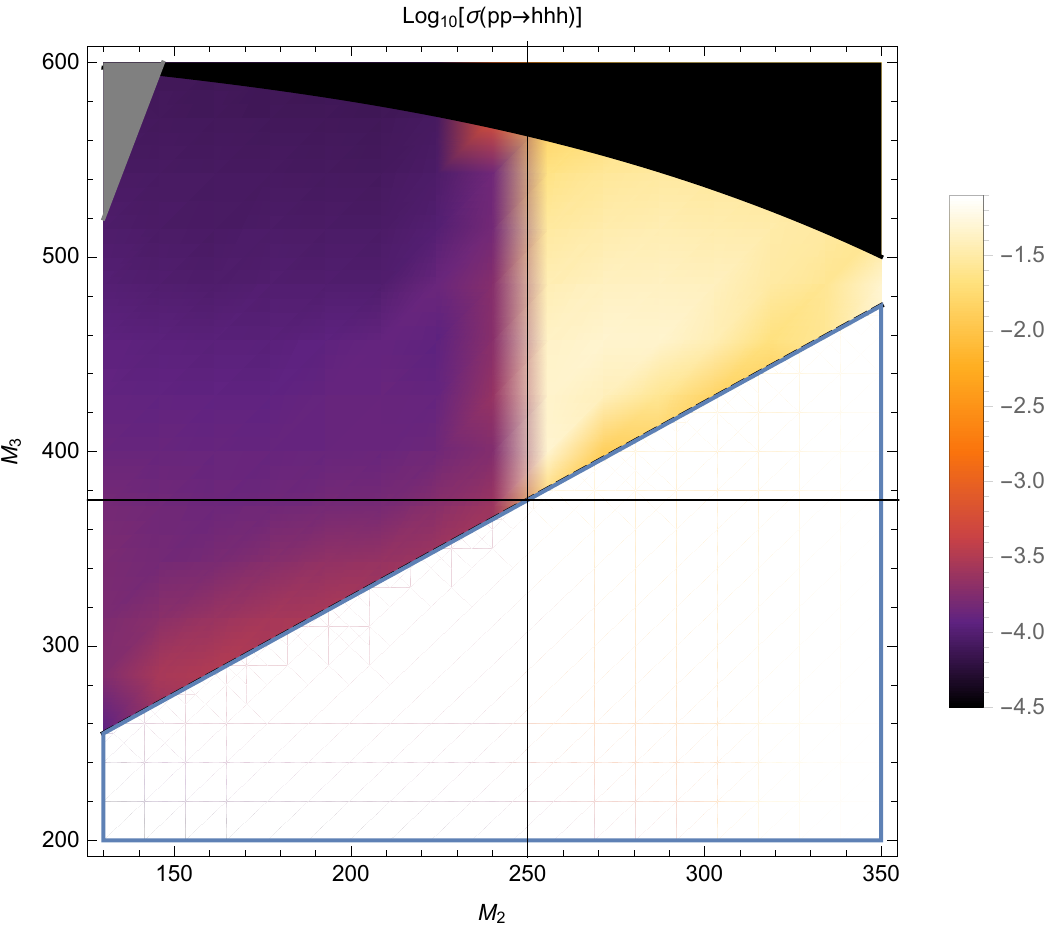}
\caption{\label{fig:hhh} Production cross sections for $h_1\,h_1\,h_1$ production in BP3 at leading order, taken from \cite{Papaefstathiou:2020lyp}. See text for more details.}
\end{center}
\end{figure}
\end{center}

In \cite{Papaefstathiou:2020lyp}, we focussed on one particular benchmark plane (BP3), that features the first production mode, in the scenario with $h_1\,\equiv\,h_{125}$. This allows for a $h_{125}\,h_{125}\,h_{125}$ final state; production cross sections depend on the masses of the two additional scalars and are displayed in figure \ref{fig:hhh}. We investigated the scenario where all $h_{125}$ further decay into $b\,\bar{b}$ final states and conducted a complete phenomenological study for a 14 \TeV~ LHC.  We made use of a customized  \texttt{loop\_sm} model implemented in \texttt{MadGraph5\_aMC@NLO} (v2.7.3)~\cite{Alwall:2014hca,Hirschi:2015iia}, that was interfaced to \texttt{HERWIG} (v7.2.1)~\cite{Bahr:2008pv,Gieseke:2011na,Arnold:2012fq,Bellm:2013hwb,Bellm:2015jjp,Bellm:2017bvx,Bellm:2019zci}. We have applied a semi-automated cut prescription to suppress SM background. Results are shown in table \ref{tab:hhh}. We see that several benchmark points are already accessible with a relatively low integrated luminosity.
{\small
\begin{table}
\begin{center}
{\small
\begin{tabular}{c||cc||cc}\\
{\bf $(M_2, M_3)$}& $\sigma(pp\rightarrow h_1 h_1 h_1)$ &
$\sigma(pp\rightarrow 3 b \bar{b})$&$\text{sig}|_{300\rm{fb}^{-1}}$& $\text{sig}|_{3000\rm{fb}^{-1}}$\\
${[\GeV]}$ & ${[\fb]}$  & ${[\fb]}$ & &\\
\hline\hline
$(255, 504)$ & $32.40$ & $6.40$&$2.92$&{  $9.23$}\\
$(263, 455)$ & $50.36$ & $9.95$&{  $4.78$}&{  $15.11 $}\\
$(287, 502)$ & $39.61$ & $7.82$&{  $4.01$} &{  $12.68$}\\
$(290, 454)$ & $49.00$ & $9.68$&{  $5.02$}&{  $15.86 $}\\
$(320, 503)$ & $35.88$& $7.09$& {  $3.76 $}&{  $11.88$}\\
$(264, 504)$ & $37.67$ & $7.44$&{  $3.56 $}&{  $11.27 $}\\
$(280, 455)$& $51.00$ & $10.07$&{  $5.18$} &{  $16.39$}\\
$(300, 475)$&$43.92$& $8.68$&{  $4.64 $}&{  $14.68 $}\\
$(310, 500)$& $37.90$ & $7.49$&{  $4.09 $}&{  $12.94$}\\
$(280, 500)$& $40.26$& $7.95$&{  $4.00 $}&{  $12.65 $}\\
\end{tabular}
}
\end{center}
\caption{\label{tab:hhh} 6 b final state {leading-order} production cross sections at 14 \TeV, as well as significances for different integrated luminosities. Taken from \cite{Papaefstathiou:2020lyp}.}
\end{table}
}
We refer the reader to the above work for details of the analysis as well as SM background simulation. Several of the benchmark points are in the 4-5 $\sigma$ range already for a relatively low luminosity, and all have significances above the discovery reach after the full run of HL-LHC. 

\begin{center}
\begin{figure}[htb]
\begin{center}
\includegraphics[width=0.48\textwidth]{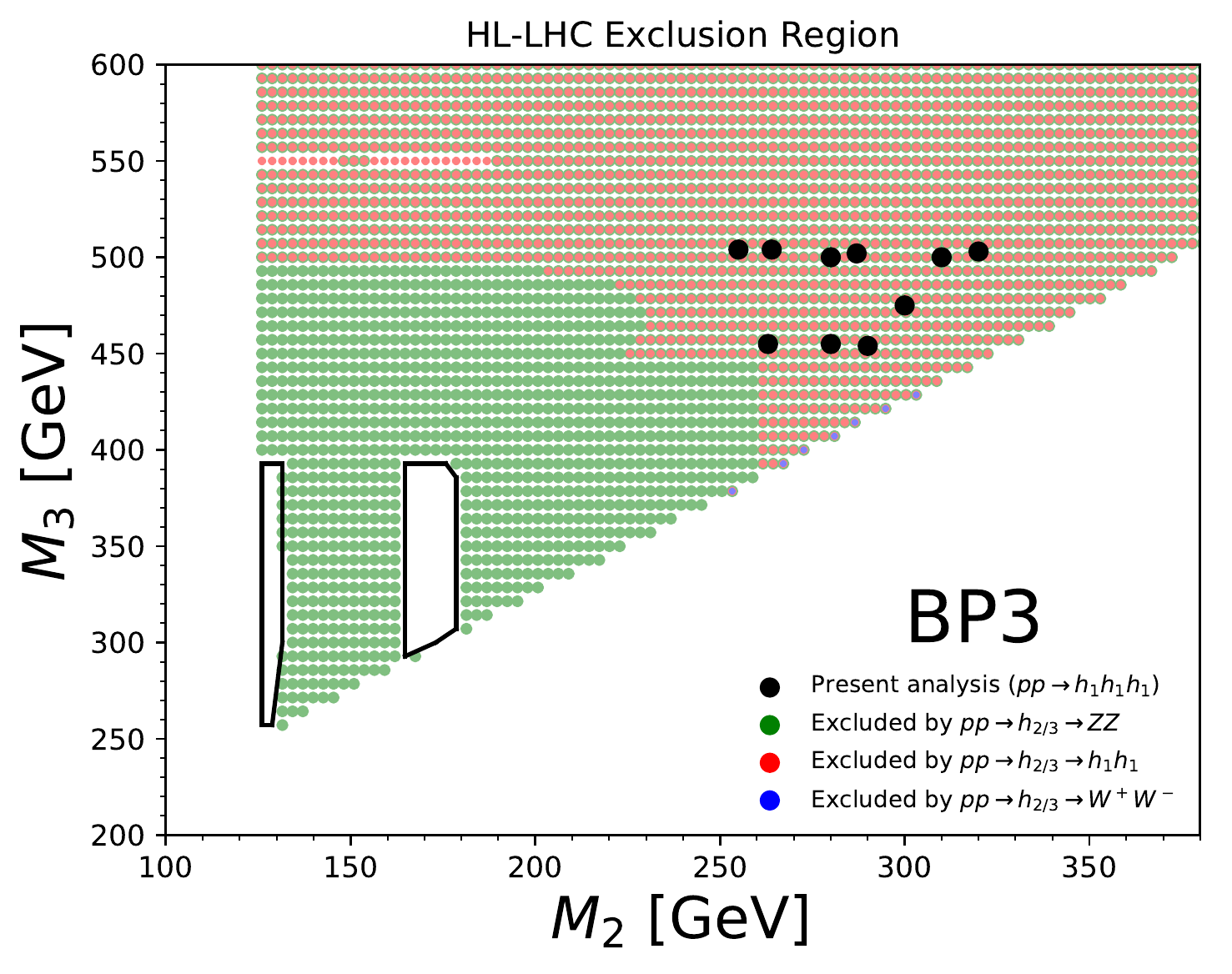}
\end{center}
\caption{Constraints on the $\lb M_2,\,M_3 \rb$ plane from extrapolation of other searches at the HL-LHC from extrapolation (see text for details). Taken from \cite{Papaefstathiou:2020lyp}.}
\label{fig:hlothers}
\end{figure}
\end{center}

Finally, we can ask whether other channels can not equally constrain the allowed parameter space at the HL-LHC. We therefore extrapolated various analyses assessing the heavy Higgs boson prospects of the HL-LHC in final states originating from $h_i \rightarrow h_1 h_1$ \cite{Sirunyan:2018two,Aad:2019uzh}, $h_i \rightarrow ZZ$ \cite{Sirunyan:2018qlb,Cepeda:2019klc} and $h_i \rightarrow W^+W^-$ \cite{Aaboud:2017gsl,ATL-PHYS-PUB-2018-022}, for $i=2,3$, and combined these with extrapolations of results from 13 TeV where appropriate. We display the results in figure \ref{fig:hlothers}.

In particular $ZZ$ final states can probe nearly all of the available parameter space. These however depend on different model parameters than the $h_1\,h_1\,h_1$ final state rates. These searches are testing a different part of the parameter space and new physics potential. 

\section{THDMa}
The THDMa has been widely promoted within the LHC Dark matter working group. It is a type II two-Higgs-doublet model that is extended by an additional pseudoscalar $a$ mixing with the "standard" pseudoscalar $A$ of the THDM. In the gauge-eigenbasis, the additional scalar serves as a portal to the dark sector, with a fermionic dark matter candidate, denoted by $\chi$. More details can e.g. be found in \cite{Ipek:2014gua,No:2015xqa,Goncalves:2016iyg,Bauer:2017ota,Tunney:2017yfp,LHCDarkMatterWorkingGroup:2018ufk,Robens:2021lov}. 

The following mass eigenstates are incorporated within this model in the scalar and dark matter sector: ${h,\,H,\,H^\pm}$, ${a,}\,{A,}\,{{\chi}}$. It depends on 12 additional new physics parameters
\begin{eqnarray*}
{v,\,m_h,\,m_H,}\,{ m_a,}\,{m_A,\,m_{H^\pm},}\,{m_\chi};\;{\cos\lb \be-\al\rb,\,\tan\be,}\,{\sin\theta;\;y_\chi,}\,{\lam_3,}\,{\lam_{P_1},\,\lam_{P_2}},
\end{eqnarray*}
where $v$ and either $m_h$ or $m_H$ are fixed by current measurements in the electroweak sector. 

\begin{center}
\begin{figure}
\begin{center}
\begin{minipage}{0.45\textwidth}
\begin{center}
\includegraphics[width=\textwidth]{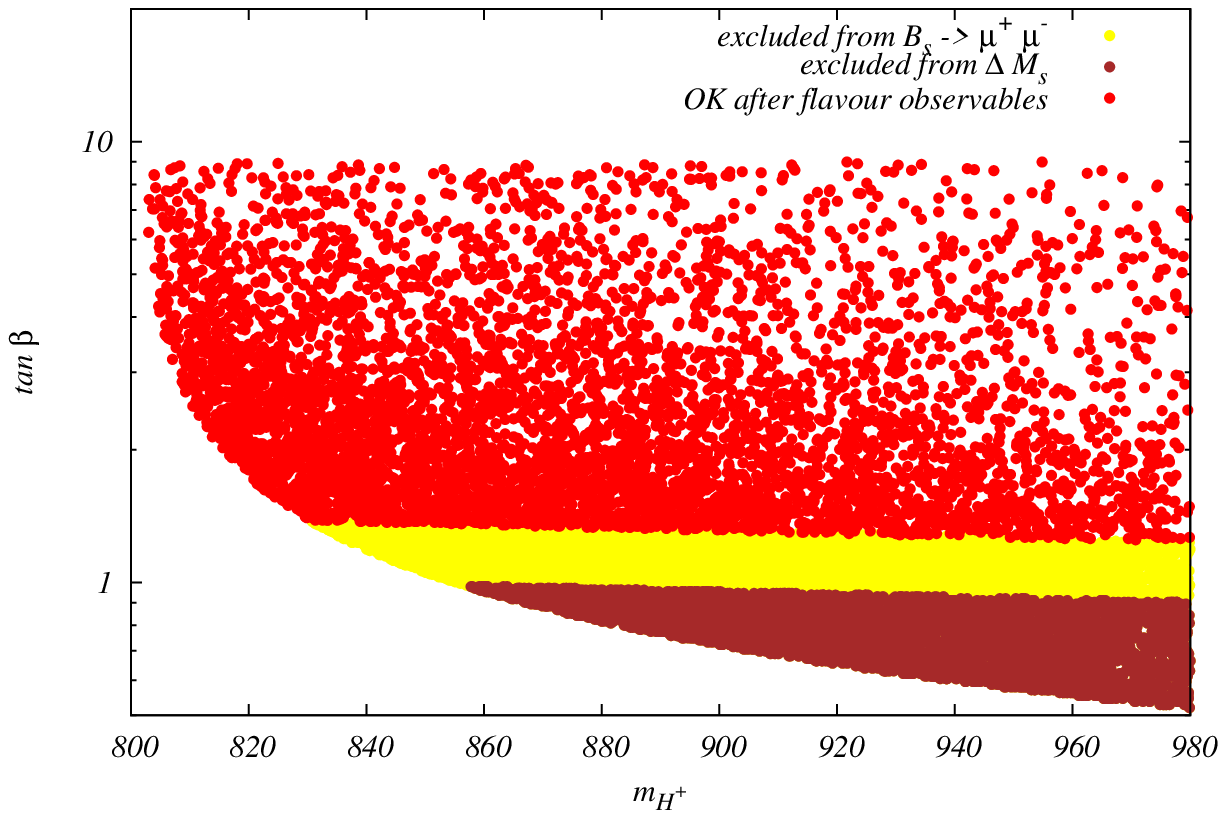}
\end{center}
\end{minipage}
\begin{minipage}{0.45\textwidth}
\begin{center}
\includegraphics[width=\textwidth]{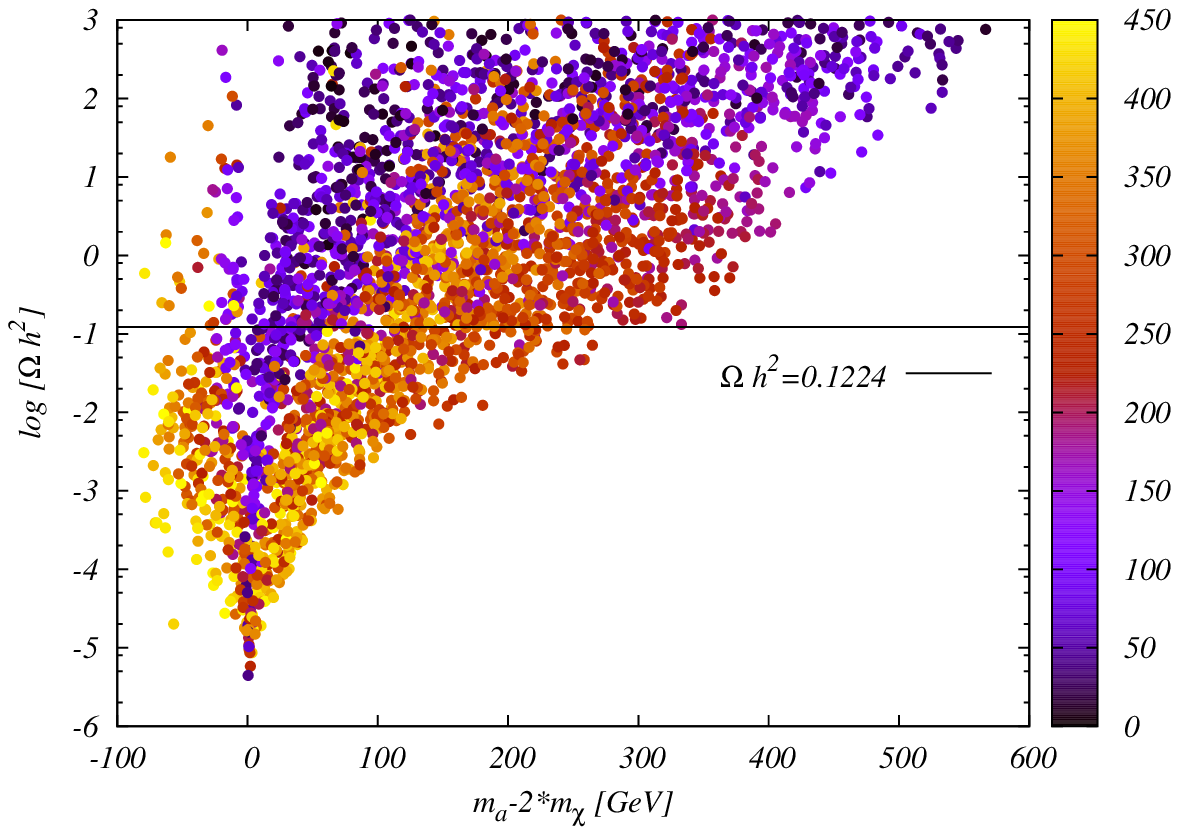}
\end{center}
\end{minipage}
\end{center}
\caption{\label{fig:thdmab} {\sl Left:} Bounds on the $\lb m_{H^\pm},\,\tan\be\rb$ plane from B-physics observables, implemented via the SPheno \cite{Porod:2011nf}/ Sarah \cite{Staub:2013tta} interface,  and compared to experimental bounds \cite{combi,Amhis:2019ckw}. The contour for low $\lb m_{H^\pm,\,\tan\be}\rb$ values stems from \cite{Misiak:2020vlo,mm}. {\sl Right:} Dark matter constraints in the THDMa model. {\sl Right:} Dark matter relic density as a function of $m_a-2\,m_\chi$, with $m_\chi$ defining the color coding. The typical resonance-enhanced relic density annihilation is clearly visible. Figures taken from \cite{Robens:2021lov}.}
\end{figure}
\end{center}

In \cite{Robens:2021lov}, a scan was presented that allows all of the above novel parameters float in specific predefined ranges. It is then not always straightforward to display bounds from specific constraints in 2-dimensional planes. Two examples where this is possible are shown in figure \ref{fig:thdmab}. The first plot displays bounds in the $\lb m_{H^\pm},\,\tan\be \rb$ plane from B-physics observables, and shows that in general low masses $m_{H^\pm}\lesssim\,800\,\GeV$  as well as values $\tan\be\lesssim\,1$ are excluded. The second plot displays the relic density as a function of the mass difference $m_a-2\,m_\chi$. In the region where this mass difference remains small, relic density annihilates sufficiently to stay below the observed relic density bound. On the other hand, too large differences lead to values $\Omega\,h_c\,\gtrsim\,0.12$ and therefore are forbidden \cite{Planck:2018vyg}.

\begin{center}
\begin{figure}
\begin{center}
\begin{minipage}{0.45\textwidth}
\begin{center}
\includegraphics[width=\textwidth]{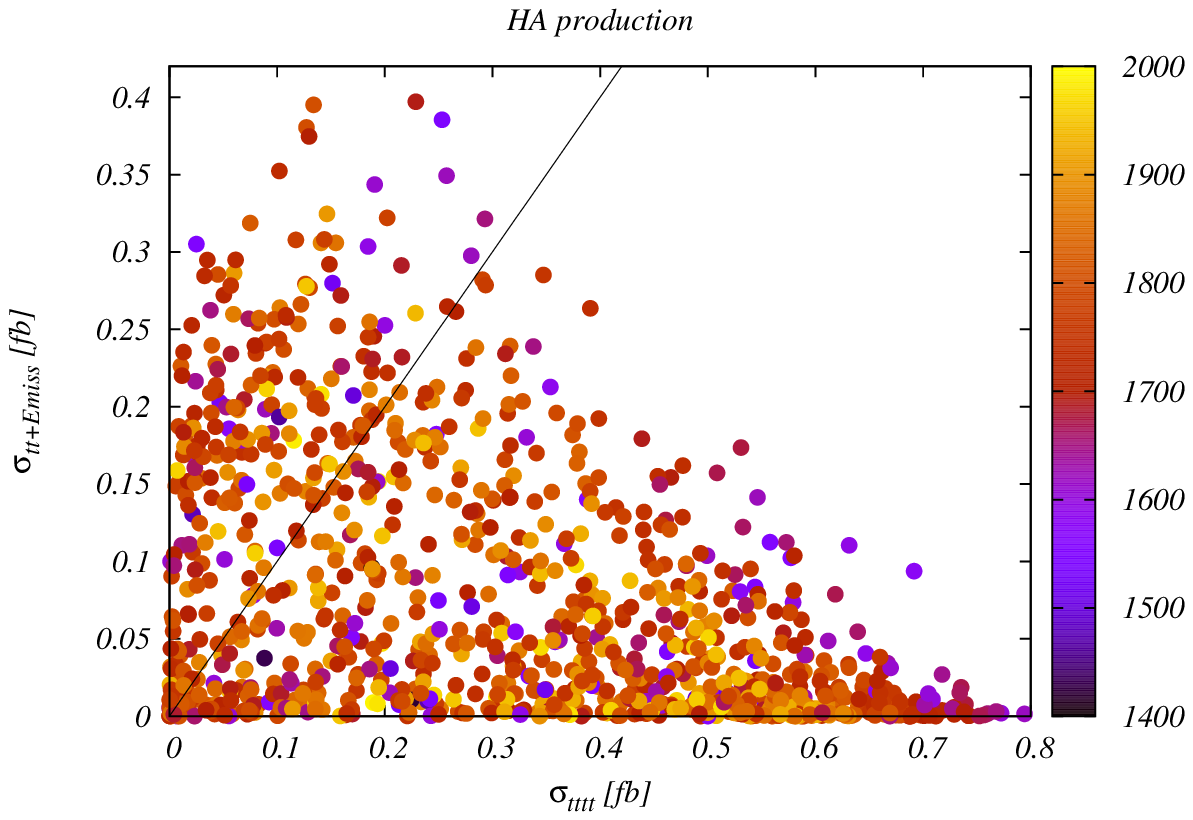}
\end{center}
\end{minipage}
\begin{minipage}{0.45\textwidth}
\begin{center}
\includegraphics[width=\textwidth]{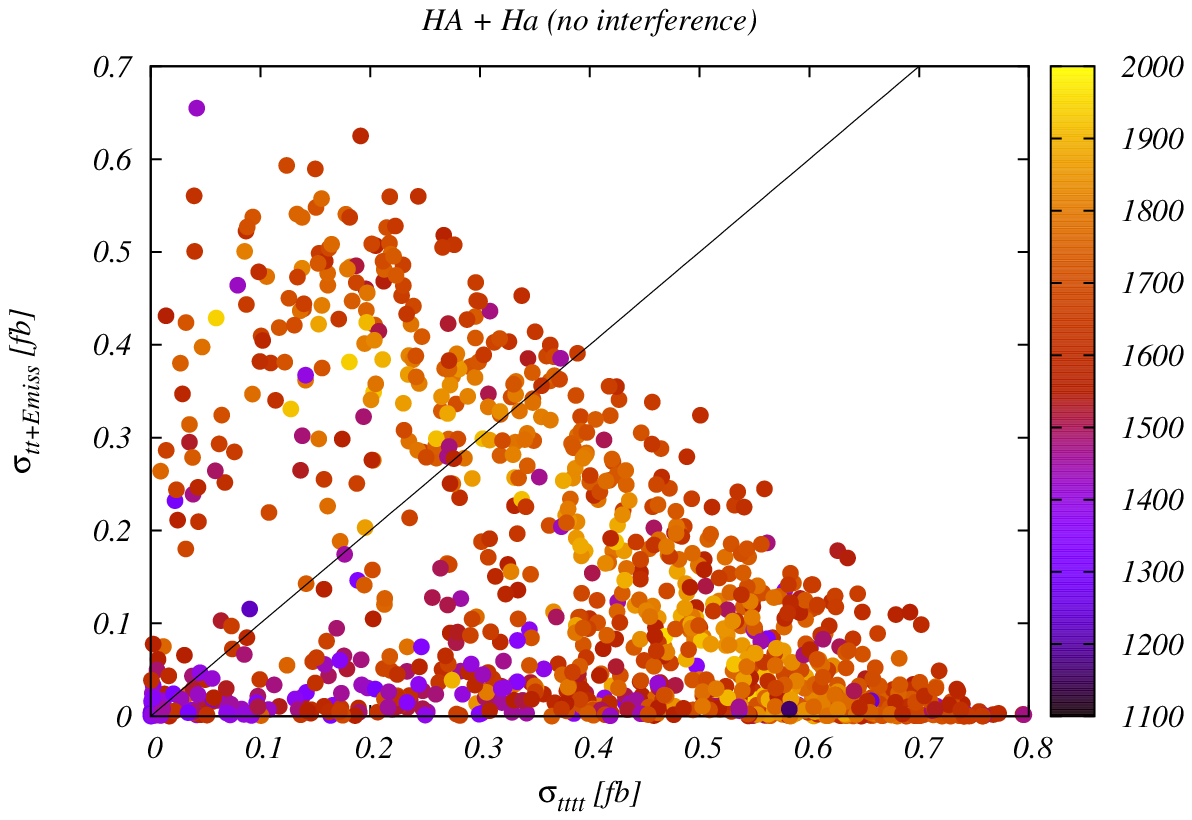}
\end{center}
\end{minipage}
\end{center}
\caption{\label{fig:thdmaatee} Production cross sections for $t\bar{t}t\bar{t}$ (x-axis) and $t\bar{t}+\slashed{E}$ (y-axis) final state in a factorized approach, for an $e^+e^-$ collider with a 3 \TeV center-of-mass energy. {\sl Left:} mediated via $HA$, {\sl right:} mediated via $HA$ and $Ha$ intermediate states. Color coding refers to $m_H+m_A$ {\sl (left)} and $M_H+0.5\times\,\lb m_A+m_a\rb$ {\sl(right)}. Figures taken from \cite{Robens:2021lov}.}
\end{figure}
\end{center}

For these scans, it was investigated which cross-section values would still be feasible for points that fulfill all constraints \cite{Robens:2021lov} at $e^+e^-$ colliders. A particular interest lies on signatures that include missing energy and therefore distinguish this model from signatures that would be realized in a THDM without a portal to the dark sector. Processes like $e^+e^-\,\rightarrow\,hA, ha$ are suppressed due to alignment, which makes $e^+e^-\,\rightarrow\,HA, Ha$ the most interesting channel that contains novel signatures. However, such parameter points typically have mass scales $\gtrsim\,1\,\TeV$. In such a case, production cross sections for an $e^+e^-$ collider with a center-of-mass energy of 3 \TeV\ are of interest. The corresponding production cross sections are shown in figure \ref{fig:thdmaatee}, which displays  predictions for $t\,\bar{t}\,t\,\bar{t}$ and $t\,\bar{t}+\slashed{E}$ final states using a factorized approach. There is a non-negligible number of points where the second channel is dominant. A "best" point with a large rate for $t\,\bar{t}+\slashed{E}_\perp$ has been presented in \cite{Robens:2021lov}.

\section{Summary and Conclusion}
In this work, I have reported on some previously published results for models that extend the scalar sector of the SM by additional gauge singlets or doublets. Some of the models discussed here in addition feature a dark matter candidate. I have presented results of applying current constraints on these models, and rendered predictions for rates or significances at various future collider options. Some of the models presented here, in particular the IDM and TRSM, have not yet been fully explored by current collider experiments. I therefore strongly encourage the experimental collaborations to consider these at LHC Run III.

\section{Acknowledgements}
This research was supported in parts by the National Science Centre, Poland, the HARMONIA project under contract UMO-2015/18/M/ST2/00518 and OPUS project under contract UMO-2017/25/B/ST2/00496 (2018-2021), by the European Union through the Programme Horizon 2020 via the COST actions CA15108 - FUNDAMENTALCONNECTIONS and CA16201 - PARTICLEFACE, and by the UK's Royal Society.
\bibliography{lit}
\end{document}